\documentclass[a4paper]{article}

\usepackage{INTERSPEECH2020}
\usepackage{subcaption}
\usepackage{graphicx}
\usepackage{lipsum}

\title{Monolingual Data Selection Analysis for English-Mandarin Hybrid Code-switching Speech Recognition}

\name{Haobo Zhang$^1$, Haihua Xu$^2$, Van Tung Pham$^2$, Hao Huang$^{1*}$, Eng Siong Chng$^{2,3}$}

\address{
  $^1$School of Information Science and Engineering, Xinjiang University, Urumqi, China\\
  $^2$Temasek Laboratories, Nanyang Technological University, Singapore \\
  $^3$School of Computer Science and Engineering, Nanyang Technological University, Singapore}
\email{\{hallboo147,hhx502,hwanghao\}@gmail.com}

\begin{document}

\maketitle
\newcommand\blfootnote[1]{%
\begingroup
\renewcommand\thefootnote{}\footnote{#1}%
\addtocounter{footnote}{-1}%
\endgroup
}

\blfootnote{*Hao Huang is the corresponding author, hwanghao@gmail.com}
\blfootnote{Work performed as an intern at Nanyang Technological University. This work was supported by the National Key R\&D Program of China (2017YFB1402101), Natural Science Foundation of China (61663044, 61761041).}

\begin{abstract}
  In this paper, we conduct data selection analysis in building an English-Mandarin code-switching (CS) speech recognition (CSSR) system, which is aimed for a real CSSR contest in China. The overall training sets have three subsets, i.e., a code-switching data set, an English (LibriSpeech) and a Mandarin data set respectively. The code-switching data are Mandarin dominated.  First of all, it is found using the overall data yields worse results, and hence data selection study is necessary. Then to exploit monolingual data, we find data matching is crucial.  Mandarin data is closely matched with the Mandarin part in the code-switching data, while English data is not. However, Mandarin data only helps on those utterances that are significantly Mandarin-dominated. Besides, there is a balance point, over which more monolingual data will divert the CSSR system, degrading results. Finally, we analyze the effectiveness of combining monolingual data to train a CSSR system with the HMM-DNN hybrid framework. The CSSR system can perform within-utterance code-switch recognition, but it still has a margin with the one trained on code-switching data.
  % As a result, More Mandarin data helps, while more English data degrades results. 
 % Secondly, data balance is also important. Merging over-imbalanced Mandarin data can result in performance drop on each individual languages. 
\end{abstract}

\section{Introduction}
Recently, code-switching speech recognition (CSSR) has drawn increased attention in automatic speech recognition (ASR) community\cite{yilmaz2018cs,guo2018cs,zeng2018e2e,Shan2019cs, khassanov2019e2e}. Here, code-switching means a linguistic phenomenon that one speaks different languages within an utterance or between consecutive utterances.
Intuitively, we can build a CSSR system simply by means of merging two languages. However, such a CSSR system can rarely produce optimal recognition results.
This is because there is no within-an-utterance code-switching samples, and the resulting ASR system frequently fails to recognize those code-switching utterances\cite{khassanov2019e2e,toshniwal2018multilingual}.  

In reality, we usually have very limited code-switching training data, but more monolingual data~\cite{khassanov2019e2e}. As a result, we treat CSSR as a low-resource ASR problem~\cite{das2015tl,Dalmia2018low-res-asr,Tung2020low-res-asr}. The question is how to fully exploit those ``unlimited" monolingual data to boost the CSSR performance.  

In this paper, we report our efforts in terms of data selection\footnote{Here, data selection simply means how to reasonably exploit monolingual data for a CSSR system with a specific code-switching pattern.}~\cite{itoh2012data-selection,wei2014data-selection-lvcsr,wei2014data-selection-unsup} for an English-Mandarin code-switching speech recognition contest sponsored by DataTang\footnote{Datatang: https://www.datatang.ai/} in China. It drew over 70+ participants worldwide. The organizer released two training data sets, namely, 200 hours of English-Mandarin code-switching data, and 500 hours of Mandarin data of which there are about 15 hours of similar code-switching data. Additionally, 960 hours of LibriSpeech English data~\cite{chen2015prn-lex,albert2018e2e-libri} is allowed to use. There are three tracks, and one of the tracks is for the hybrid DNN-HMM based ASR system competition. Except for the above-mentioned three training data sets,  a trigram language model is given as well,  and the only flexibility is the participants are allowed to use their own lexicon to build their ASR system. In this work, all the ASR results are based on the DNN-HMM system~\cite{povey2018tdnnf}.

The motivation of this work lies in the observation that we end up with worse results when using the overall three data sets. This motivates us to answer how to exploit the two monolingual data sets reasonably. To begin with, we first add Mandarin data incrementally, and we found consistent performance improvement on the Mandarin part recognition, with insignificant performance drop on the English part. Nevertheless, we found that there is a limit point, where more data degrade results over the point. We do the same experiments for LibriSpeech English data. however, we observe consistent performance drop. We conjecture this is due to the data mismatch problem. To look further into the problem,
we categorize the test utterances into different subsets according to how many 
English words each utterance contains, then analyze the performance change on those subsets respectively.
% The Mandarin data are closely matched with the corresponding Mandarin part of the code-switching data, while LibriSpeech English data is to the contrary. 
% We verfiy

% We have no matched English training data to verify our conjecture, However, we have mismatched SEAME test data \cite{}, which is
% Malaysia-Singapore English-Mandarin code-switching data. It is significantly mismatched with the training data here for either English or Mandarin. Therefore,
% we verify the conjecture by employing the two ASR systems to recognize the SEAME code-switching test data\cite{}. Though mismatched, we observe the effectiveness by using more data from either language.

Moreover, we are interested in a condition where no code-switching data is available at all.
As a result, one has to utilize monolingual data to train a code-switching system.
Particularly, we are interested in how much effectiveness can be achieved by merging pure monolingual data, compared with that we have code-switching data. 
We remove all the code-switching data from the 500 Mandarin data set, and incrementally merge the LibriSpeech English data into it to train a CSSR system. 
To our surprise, using more English data is consistently beneficial to both language performance improvement. We notice that this is one of the significant differences of the DNN-HMM ASR system versus End-to-End ASR~\cite{zhou2018comparison-e2e,karita2019comparative-e2e} one. By merging monolingual data, it is observed 
end-to-end CSSR system fails to recognize those utterances containing code-switching words~\cite{khassanov2019e2e,toshniwal2018multilingual}.

The paper is organized as follows. Section~\ref{sec:dd} describes the overall data for the experiments that are followed. Section~\ref{sec:e-setup} reports the experimental setup.
Section~\ref{sec:baseline} describes the development of our baseline system. Section~\ref{sec:add-mono} analyzes the effectiveness of using more Mandarin monolingual data, as well as more English data. 
% Section \ref{sec:mismatch} analyzes mismatching problem, and 
Section~\ref{sec:pure-mono} reports the effectiveness of using monolingual data to train
CSSR system. Finally we concludes in Section~\ref{sec:con}.
% This template can be found on the conference website. Templates are provided for Microsoft Word\textregistered, and \LaTeX. However, we highly recommend using \LaTeX when preparing your submission. Information for full paper submission is available on the conference website.

\section{Data description} \label{sec:dd}
The entire data sets that are released by the organizer are reported in Table \ref{tab:dataset}.
Each participant has three training data sets and three test data sets, two \textit{dev} sets, and 
one evaluation set that is released after the contest. 

\begin{figure}[th]
\centering
\captionsetup{justification=centering}
\includegraphics[width=6cm]{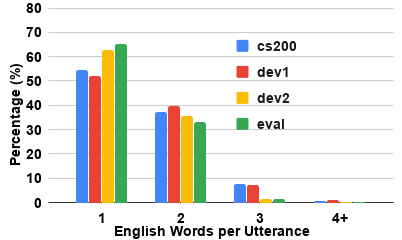}
\caption[]{English word distribution of the code-switching \\data sets, including T1, E1 to E3 in Table ~\ref{tab:dataset}.}
\label{fig:cs-hist}
\end{figure}

The code-switching data are  Mandarin-dominated. Figure~\ref{fig:cs-hist}
reveals the English word distribution in  T1, and E1 to E3 four
code-switching data sets in Table~\ref{tab:dataset}. As is shown in Figure~\ref{fig:cs-hist}, the majority of utterances contain only
single English words. Besides, English word distributions are very similar between the training data T1 and the \textit{dev} set E1, while the cases of E2 and E3 are more alike.
\begin{table}[th]
  \caption{Data sets and length description}
  \label{tab:dataset}
  \centering
  \begin{tabular}{c c c}
    \toprule
    \multicolumn{1}{c}{\textbf{Notation}} & \multicolumn{1}{c}{\textbf{Data\ Set}} & \multicolumn{1}{c}{\textbf{Hours}}\\
    \midrule
    $T1$ & $\text{CS200}$ & $200$~~~\\
    $T2$ & $\text{Man500}$ & $500$~~~\\
  %  $T3$ & $\text{cs15}$ & $15$~~~\\
    $T3$ & $\text{Libri}$ &  $960$~~~\\
    \midrule
    $E1$ & $\text{dev1}$ & $20$~~~\\
    $E2$ & $\text{dev2}$ & $20$~~~\\
    $E3$ & $\text{eval}$ & $20$~~~\\
    \bottomrule
  \end{tabular}
\end{table}

\section{Experimental setup}\label{sec:e-setup}
All experiments are conducted with Kaldi\footnote{https://kaldi-asr.org/}.
The acoustic models are trained with the Lattice-free Maximum Mutual Information (LF-MMI) criterion~\cite{povey2016lf-mmi} over the Factorized Time-delay Neural Network (TDNNf)~\cite{povey2018tdnnf}.
The acoustic feature is the concatenation of 40-dimensional MFCC features and 100-dimensional i-vectors~\cite{saon2013ivector,peddinti2015tdnn}. 
% The model in experiments of add monolingual data and pure monolingual data 
The TDNNf is made up of 15 layers, and each layer is decomposed as 1536x256, 256x1536, where 256 is the dimension of the bottleneck layer. Besides, the activation function is 
Rectified Linear Unit (ReLU)~\cite{dahl2013relu}. To train TDNNf, data augmentation is employed~\cite{ko2015da}.

The vocabulary is $\sim$200k, of which Mandarin and English words are $\sim$121k and $\sim$79k respectively.
We use language-dependent phone set, and there are 210 Mandarin initials and finals~\cite{guo2018cs}, as well as 42 English phones.
As mentioned, the language models are released by the organizer, and it is a trigram language model that includes $\sim$813k Mandarin words and $\sim$116k English words.

%%981996
\section{Baseline} \label{sec:baseline}
We report our efforts on building the baseline system.
Since the data and language models are fixed, our efforts are mainly focused on choosing an appropriate lexicon. Initially, we use our in-house lexicon. As there is a mismatch between the language models and the lexicon in Chinese
word segmentation, we decompose all out-of-vocabulary Chinese words into characters, yielding our initial lexicon, denoted as L0 here. After submitting our evaluation results, we realized there are many entries with incorrect pronunciations for the Chinese part in L0. We conducted manual checking and updated L0 to L1.
% Finally, we extracted the overall words from the organizer language models, and 
% labelled the pronunciations for them, and we got lexicon L2.
Simultaneously, we also attempt to use a grapheme lexicon for English words, and this is inspired by \cite{duc2019grapheme}.
Table \ref{tab:baseline-lex} reports our baseline results in terms of Meta-data Error Rate (MER) using only \textit{CS200} code-switching training data. By MER, we mean a token error rate that is a mixture of Chinese character error rate and English word error rate.

\begin{table}[th]
  \caption{MER (\%) with different efforts on the recognition lexicons; the systems are trained with \textit{CS200} code-switching training data}
  \label{tab:baseline-lex}
  \centering
  \begin{tabular}{c c c c}
    \toprule
    \multicolumn{1}{c}{\textbf{Dictionary}} & \multicolumn{3}{c}{\textbf{MER (\%)}}\\
    $ $ & dev1 & dev2 & eval \\
    \midrule
    L0 & $7.43$ & $6.81$ & $7.51$~~~\\
    L1 & $7.07$ & $6.28$ & $7.08$~~~\\
    Grapheme &  $7.11$ & $6.4$ & $7.2$  ~~~\\
    % $L2$ &  $6.75$ & $5.97$ & $6.79$  ~~~\\
    \bottomrule
  \end{tabular}
\end{table}
From Table~\ref{tab:baseline-lex}, The lexicon after manual checking gets the best results. Therefore, we use L1 for the remaining experiments in this paper. % We notice that even our L1 Lexicon is mismatched with the  language models. We realized this after result submission. Recently, we resolved the inconsistency issue, and obtained better results.
Table~\ref{tab:baseline-overall} reports experimental results using various data combination recipe.

\begin{table}[th]
  \caption{MER (\%) results by combining various training data set to train the ASR system}
  \label{tab:baseline-overall}
  \centering
  \begin{tabular}{l c c c}
    \toprule
    \multicolumn{1}{c}{\textbf{Data}} & \multicolumn{3}{c}{\textbf{MER (\%)}}\\
    $ $ & dev1 & dev2 & eval \\
    \midrule
    %$cs200$ & $7.58$ & $6.58$ & $7.39$\\
    CS200 & $7.07$ & $6.28$ & $7.08$\\
    +CS15 & $6.93$ & $6.14$ & $6.87$\\
    +Man500 & $\textbf{6.87}$ & $\textbf{5.91}$ & $\textbf{6.63}$\\
    +Libri & $7.58$ & $6.67$ & $7.50$\\
    \bottomrule
  \end{tabular}
\end{table}
Table~\ref{tab:baseline-overall} suggests employing more Mandarin data works, yielding significant MER reduction, however, employing more \textit{Libri} English data degrades 
results. These suggest there are data selection issues concerned,
and it is worthwhile to see what details look like.

% Intuitively, These might be involved with data mismatch
% problem, or even data balance problem.  More importantly,
% more details are yet to be unveiled. To begin with, we are interested to see the effects of adding each  monolingual data 
% individually.

\section{Analysis of adding monolingual data} \label{sec:add-mono}
\subsection{Mandarin data} \label{sub:add-mandarin}
\begin{figure} [th]
    \centering
    \quad
    \begin{subfigure}[b]{0.214\textwidth}
        \centering
        \includegraphics[width=\textwidth]{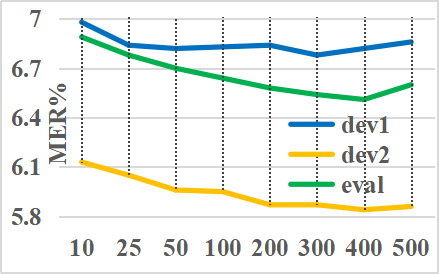}
        \caption[]%
        
    \end{subfigure}
    \quad
    \begin{subfigure}[b]{0.214\textwidth}
        \centering
        \includegraphics[width=\textwidth]{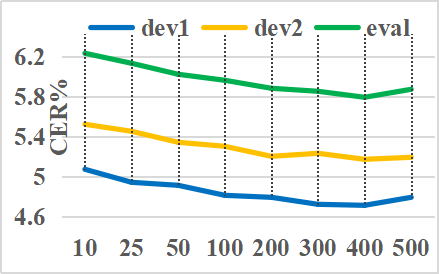}
        \caption[]%
        
    \end{subfigure}
    \vskip 0cm \quad
    \begin{subfigure}[b]{0.214\textwidth}   
        \centering
        \includegraphics[width=\textwidth]{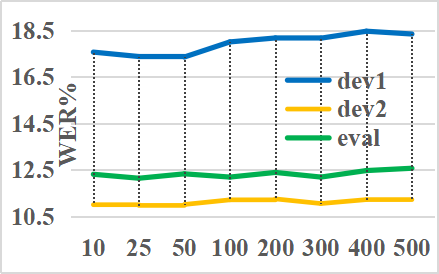}
        \caption[]%

    \end{subfigure}
    \quad
    \begin{subfigure}[b]{0.214\textwidth}   
        \centering
        \includegraphics[width=\textwidth]{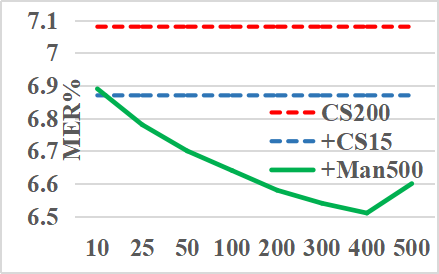}
        \caption[]%

    \end{subfigure}
    \caption[]
    {Recognition results change versus Mandarin data Selection from \textit{Man500} data set. (a) Overall MER (\%) change on the \textit{dev1}, \textit{dev2}, and \textit{eval} three data sets. (b) Character error rate (CER) (\%) for the Mandarin part in the three test sets. (c) English WER (\%) in the three test sets. (d) \textit {eval} MER (\%) comparison between the systems with different training sets.}
    \label{fig:mandarin-data}
\end{figure}

To begin with, we fix the \textit{CS200} code-switching data, incrementally increasing monolingual Mandarin data from the \textit{Man500}. We are interested to see how the recognition results will be changed for each individual language, as well as for the two languages combined. Figure~\ref{fig:mandarin-data}
plots the various change of the recognition results versus the incremental increase
of Mandarin data usage. We notice that we select the data as follows. We first ensure each selection covers the overall speakers, and then determine how many utterances are to be selected, and finalize by random selection.

From Figure~\ref{fig:mandarin-data}, 
what we observe can be briefly summarized in the following aspects.
First, employing more Mandarin data generally produces improved results on the
Mandarin part as shown in Figure~\ref{fig:mandarin-data}(b).
Secondly, more Mandarin data does not significantly affects English recognition results, as indicated in Figure~\ref{fig:mandarin-data}(c). 
Thirdly, more Mandarin data over a certain point can hurt the overall performance, as can be seen in Figure~\ref{fig:mandarin-data}(d). The best performance is achieved when about 400 hours of Mandarin data is employed on the \textit{eval} test set. 

First of all, we attribute the improvement to the close similarity of the \textit{Man500} and the \textit{CS200} code-switching data in terms of the Mandarin part. Actually, they are all read speech released by the organizer.
After that, we think there is a balance point, and more monolingual Mandarin data can divert the models in terms of code-switching capability. However, 
the above-mentioned guesses need further support from more details.

\begin{figure}[th]
    \quad
    \begin{subfigure}[b]{0.214\textwidth}
        \centering
        \includegraphics[width=\textwidth]{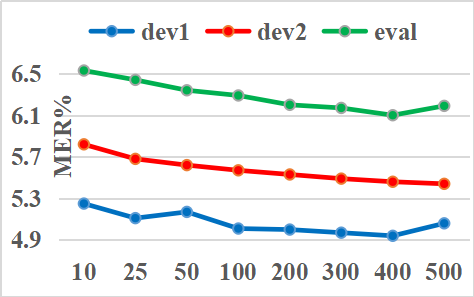}
        \caption[]%
        {{\small}}

    \end{subfigure}
    \quad
    \begin{subfigure}[b]{0.214\textwidth}  
        \centering
        \includegraphics[width=\textwidth]{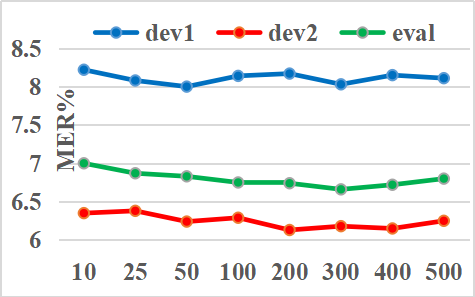}
        \caption[]%
        {{\small}}

    \end{subfigure}
    \vskip 0cm \quad
    \begin{subfigure}[b]{0.214\textwidth}
        \centering 
        \includegraphics[width=\textwidth]{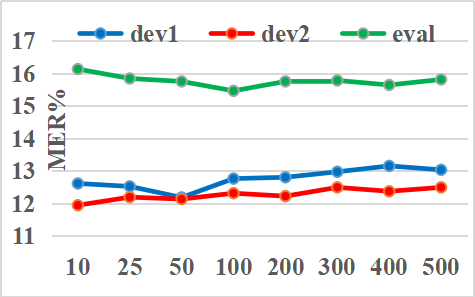}
        \caption[]%
        {{\small }}    

    \end{subfigure}
    \quad
    \begin{subfigure}[b]{0.214\textwidth}
        \centering 
        \includegraphics[width=\textwidth]{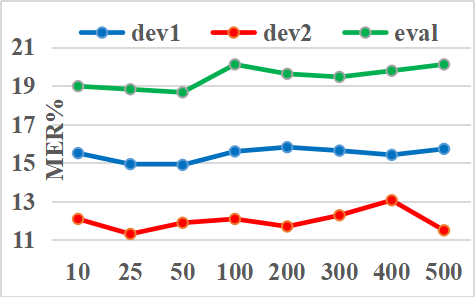}
        \caption[]%
        {{\small}}    

    \end{subfigure}
    \caption[]
    { MER (\%) change versus Mandarin data selection from \textit{Man500} data set for 4 categories of utterances as shown in Figure~\ref{fig:cs-hist}. (a) utterances with single English word; (b) utterances with 2 English words; (c) utterances with 3 English words; (d) utterances with no fewer than 4 English words.} \label{fig:mandarin-data-detail}
\end{figure}

To see how the added Mandarin monolingual data affects the results of different category of utterances as shown in Figure~\ref{fig:cs-hist}, we draw Figure~\ref{fig:mandarin-data-detail} revealing the details.
From the figure, we can see the more English words the less performance gain from more Mandarin monolingual data usage in terms of recognition results. This is particularly true for
Figure~\ref{fig:mandarin-data-detail}(c)-(d), where utterances contain no fewer than 3 English words, and the performance are not stably increased. From
Figure~\ref{fig:mandarin-data-detail}(d), using more Mandarin data actually
degrades the results on \textit{eval} test set.

Combined with what is shown in Figures~\ref{fig:mandarin-data} and \ref{fig:mandarin-data-detail},
the recognition performance gains are mainly from the utterances with the single English word. From this perspective, data matching guesses are supported here.
However,
it does limited or even negative help on the performance improvement for those utterances containing more English words. We achieved overall performance improvement as shown in Figure~\ref{fig:mandarin-data}, only because the single English utterances are dominated, as shown in Figure~\ref{fig:cs-hist}.

\subsection{English data}\label{sub:engglish-data}
In this section, we are interested to see how the English-Mandarin code-switching ASR system is affected by using more English monolingual data. Here, the English data is \textit{LibriSpeech}, as shown in Table~\ref{tab:dataset}. It is worth a mention that we 
do this over the ASR system that is trained with \textit{CS200} plus \textit{Man500}, instead of using the data by directly combining
\textit{CS200} and \textit{LibriSpeech}.

\begin{figure} [th]
    \quad
    \begin{subfigure}[b]{0.214\textwidth}
        \centering
        \includegraphics[width=\textwidth]{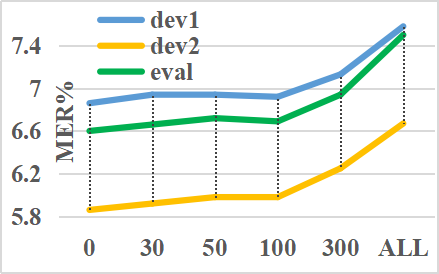}
        \caption[]%

    \end{subfigure}
    \quad
    \begin{subfigure}[b]{0.214\textwidth}  
        \centering
        \includegraphics[width=\textwidth]{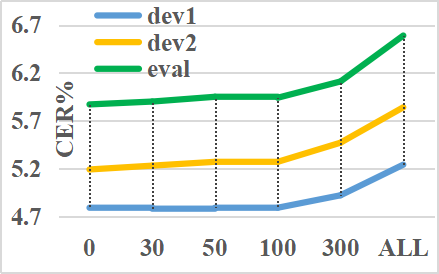}
        \caption[]%

    \end{subfigure}
    \vskip 0cm \quad
    \begin{subfigure}[b]{0.214\textwidth}
        \centering 
        \includegraphics[width=\textwidth]{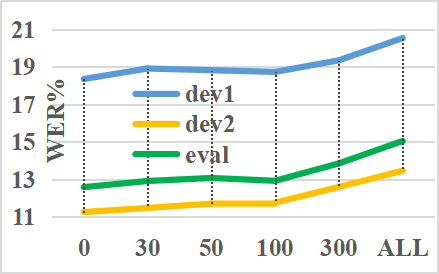}
        \caption[]%

    \end{subfigure}
    \quad
    \begin{subfigure}[b]{0.214\textwidth}   
        \centering
        \includegraphics[width=\textwidth]{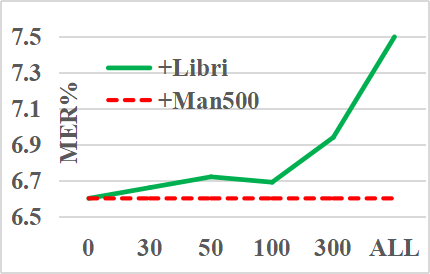}
        \caption[]%

    \end{subfigure}
    \caption[]
    {Results change versus incrementally increasing English data from \textit{LibriSpeech}. (a) Overall MER (\%) change on the three test sets; (b)\&(c)  Performance changes for Mandarin(CER\%) and English(WER\%) respectively; (d) \textit{eval} MER (\%) comparison over the system without English data employed from \textit{LibriSpeech} at all.}\label{fig:librispeech}
\end{figure}

Figure~\ref{fig:librispeech} plots the recognition performance change versus incrementally introducing \textit{LibriSpeech} data. Figure~\ref{fig:librispeech} clearly reveals recognition performance degradation as more English data are merged, on either Mandarin or English part of the code-switching utterances. Intuitively, we attribute this to the data mismatching problem. That is, \textit{LibriSpeech} is mismatched with the English that are spoken by Chinese Mainland speakers, and such a mismatch can be enlarged for the case when those single English words are mixed with Chinese words. To verify our guess, we also need further details.

\begin{figure} [th]
    \quad
    \begin{subfigure}[b]{0.214\textwidth}
        \centering
        \includegraphics[width=\textwidth]{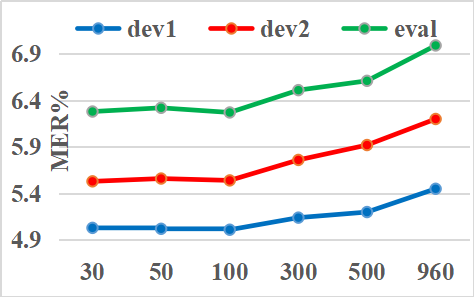}
        \caption[]%

    \end{subfigure}
    \quad
    \begin{subfigure}[b]{0.214\textwidth}  
        \centering
        \includegraphics[width=\textwidth]{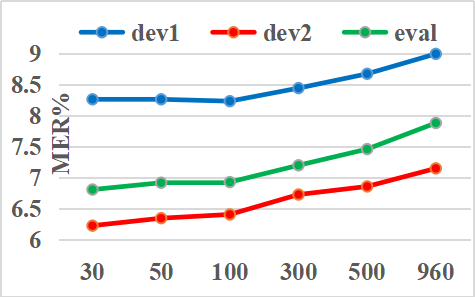}
        \caption[]%

    \end{subfigure}
    \vskip 0cm \quad
    \begin{subfigure}[b]{0.214\textwidth}
        \centering 
        \includegraphics[width=\textwidth]{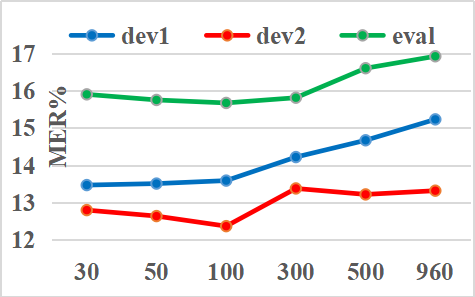}
        \caption[]%

    \end{subfigure}
    \quad
    \begin{subfigure}[b]{0.214\textwidth}
        \centering 
        \includegraphics[width=\textwidth]{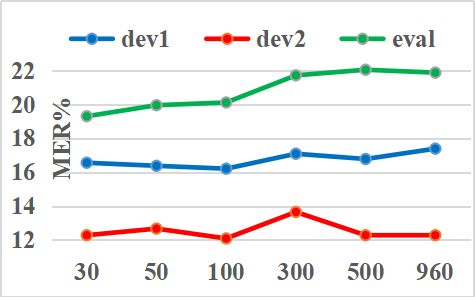}
        \caption[]%

    \end{subfigure}
    \caption[]
    {MER (\%) change versus English data selection from \textit{LibriSpeech} for 4 groups of utterances as shown in Figure~\ref{fig:cs-hist}. (a) utterances with single English word; (b) utterances with 2 English words; (c) utterances with 3 English words; (d) utterances with no fewer than 4 English words.}\label{fig:libri-detail}
\end{figure}

Figure~\ref{fig:libri-detail} shows the  performance change details on 4 groups of utterances as shown in Figure~\ref{fig:cs-hist}. From what Figure~\ref{fig:libri-detail} shows, we cannot simply attribute performance degradation only to data mismatching reason. With Mandarin dominated utterances as seen in Figure~\ref{fig:libri-detail}(a)-(b), introducing English data does not help, and more English data degrades results. However, for utterance less dominated with Mandarin words as shown in Figure~\ref{fig:libri-detail}(c)-(d), where there are more than 3 English words in the utterances, using more English data does not necessarily lead to a performance drop. This suggests data mismatch is one reason, code-switching pattern is another reason. Different code-switching patterns are affected by monolingual data usage differently. However, we really need English data from Mainland Chinese to confirm a solid conclusion.
%%%%%%%%%%%

% \section{Mismatching analysis} \label{sec:mismatch}
\section{Pure monolingual data} \label{sec:pure-mono}
\begin{figure}[t]
    \quad
    \begin{subfigure}[b]{0.214\textwidth}
        \centering
        \includegraphics[width=\textwidth]{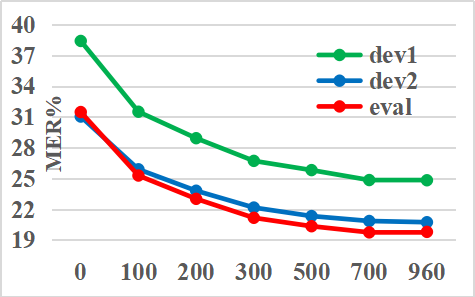}
        \caption[]%
        
    \end{subfigure}
    \quad
    \begin{subfigure}[b]{0.214\textwidth}  
        \centering
        \includegraphics[width=\textwidth]{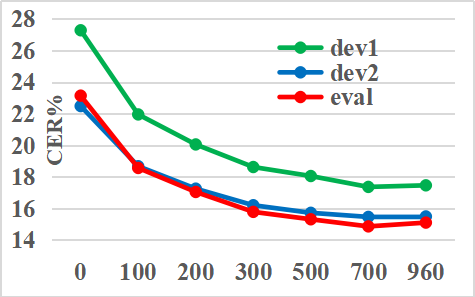}
        \caption[]%
        
    \end{subfigure}
    \vskip 0cm \quad
    \begin{subfigure}[b]{0.214\textwidth}
        \centering 
        \includegraphics[width=\textwidth]{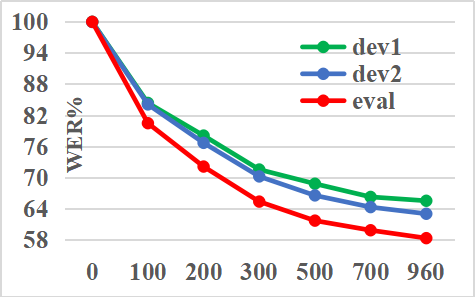}
        \caption[]%
        
    \end{subfigure}
    \quad
    \begin{subfigure}[b]{0.214\textwidth}
        \centering 
        \includegraphics[width=\textwidth]{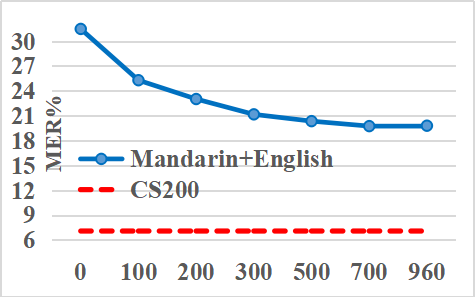}
        \caption[]%
        
    \end{subfigure}
    \caption[]
    {\small Result analysis of code-switching system built with monolingual data mergence. (a) MER (\%) on the 3 test sets; (b) CER (\%) for the Mandarin part; (c) WER (\%) for the English part; (d) \textit{eval} MER (\%) comparison between the two systems with different ``code-switching" training data.  }\label{fig:monolingual}
\end{figure}

Code-switching ASR system is much more flexible compared with monolingual ASR one. However, real code-switching data is a low-resource, and hard to access. One natural consideration is to build a code-switching ASR system by merging related monolingual data.
This has been extensively studied under the End-to-end (E2E) ASR framework\cite{duc2019grapheme}. 
Unfortunately, the state-of-the-art E2E ASR system almost completely fails to recognize those utterances mixed with word from different language (within-utterance code-switching)~\cite{duc2019grapheme}, when it is learned with monolingual data. This is because the E2E system has never been learned with utterances containing code-switching word sequence, and so does the decoder fail to predict such a sequence.
Here, we examine if such a disastrous case can happen under the HMM-DNN ASR framework. Specifically, we do as follows. We 
first make a subset of \textit{Man500} set, by
removing all code-switching utterances from it. 
We fix the Mandarin subset and keep incrementally merging the English data from \textit{LibriSpeech}.
Figure~\ref{fig:monolingual} draws recognition results with a code-switching ASR system trained with monolingual data. 
% monolingual analyze

From Figure~\ref{fig:monolingual}, we see consistent performance improvement by merging the two monolingual data sets. By looking into Figure~\ref{fig:monolingual}(b), the CER drops for the Mandarin part means the coexistence relationship of the individual language is critical in code-switching scenario. Improving one language can benefit another language's recognition performance. However, there is a balance point, and it should be related to a specific code-switching pattern as shown in Section~\ref{sec:add-mono}.
By looking into Figure~\ref{fig:monolingual}(c), 
we see sharp WER drops for the English part. However, the mismatching guess in Section~\ref{sub:engglish-data} is observable here. The WER drop ``stops" over 50\%, which is much higher than what is shown in Figure~\ref{fig:mandarin-data}(c), though the overall WER decrease has not been fully saturated.
Finally in Figure~\ref{fig:monolingual}(d), we compare the performance of the two systems on \textit{eval} set, one trained with monolingual data (though English part is not matched), and another trained with \textit{CS200} code-switching data. The gap is still remarkable, and this should be further studied under a fair comparison scenario.

\section{Conclusion} \label{sec:con}
In this work, we conducted a thorough study on the code-switching system performance with diversified data selections. 
First, we analyzed the code-switching pattern, it is Mandarin-dominated code-switching data sets, both for training and testing.
We also found the monolingual data helps, but data matching is crucial. Besides, there is a balance point for how much monolingual is employed, which is dependent on a specific code-switching environment.
Finally, we performed an analysis on a code-switching system performance assuming no real code-switching data available, under the HMM-DNN modeling framework. We found the HMM-DNN ASR system still well performs for within-utterance code-switching recognition. Our observation is 
completely different from what have been observed under the End-to-end ASR framework.

\section{Acknowledgements} \label{sec:ack}
 The computational work for this paper is partially performed on the resources of the National Supercomputing Centre (NSCC), Singapore (https://www.nscc.sg).

%\newpage
% \printbibliography
\bibliographystyle{IEEEtran}
\bibliography{mybib}

\begin{thebibliography}{10}
\providecommand{\url}[1]{#1}
\csname url@samestyle\endcsname
\providecommand{\newblock}{\relax}
\providecommand{\bibinfo}[2]{#2}
\providecommand{\BIBentrySTDinterwordspacing}{\spaceskip=0pt\relax}
\providecommand{\BIBentryALTinterwordstretchfactor}{4}
\providecommand{\BIBentryALTinterwordspacing}{\spaceskip=\fontdimen2\font plus
\BIBentryALTinterwordstretchfactor\fontdimen3\font minus
  \fontdimen4\font\relax}
\providecommand{\BIBforeignlanguage}[2]{{%
\expandafter\ifx\csname l@#1\endcsname\relax
\typeout{** WARNING: IEEEtran.bst: No hyphenation pattern has been}%
\typeout{** loaded for the language `#1'. Using the pattern for}%
\typeout{** the default language instead.}%
\else
\language=\csname l@#1\endcsname
\fi
#2}}
\providecommand{\BIBdecl}{\relax}
\BIBdecl

\bibitem{yilmaz2018cs}
E.~Y{\i}lmaz, H.~v.~d. Heuvel, and D.~A. van Leeuwen, ``Acoustic and textual
  data augmentation for improved asr of code-switching speech,'' \emph{arXiv
  preprint arXiv:1807.10945}, 2018.

\bibitem{guo2018cs}
P.~Guo, H.~Xu, L.~Xie, and E.~S. Chng, ``Study of semi-supervised approaches to
  improving {English-Mandarin} code-switching speech recognition,'' \emph{arXiv
  preprint arXiv:1806.06200}, 2018.

\bibitem{zeng2018e2e}
Z.~Zeng, Y.~Khassanov, V.~T. Pham, H.~Xu, E.~S. Chng, and H.~Li, ``On the
  end-to-end solution to {Mandarin-English} code-switching speech
  recognition,'' in \emph{20th Annual Conference of the International Speech
  Communication Association, INTERSPEECH}, 2019, in-press.

\bibitem{Shan2019cs}
C.~{Shan}, C.~{Weng}, G.~{Wang}, D.~{Su}, M.~{Luo}, D.~{Yu}, and L.~{Xie},
  ``Investigating end-to-end speech recognition for {Mandarin-English}
  code-switching,'' in \emph{ICASSP 2019 - 2019 IEEE International Conference
  on Acoustics, Speech and Signal Processing (ICASSP)}, May 2019, pp.
  6056--6060.

\bibitem{khassanov2019e2e}
Y.~Khassanov, H.~Xu, V.~T. Pham, Z.~Zeng, E.~S. Chng, C.~Ni, and B.~Ma,
  ``Constrained output embeddings for end-to-end code-switching speech
  recognition with only monolingual data,'' in \emph{20th Annual Conference of
  the International Speech Communication Association, INTERSPEECH}, 2019,
  in-press.

\bibitem{toshniwal2018multilingual}
S.~Toshniwal, T.~N. Sainath, R.~J. Weiss, B.~Li, P.~Moreno, E.~Weinstein, and
  K.~Rao, ``Multilingual speech recognition with a single end-to-end model,''
  in \emph{2018 IEEE International Conference on Acoustics, Speech and Signal
  Processing (ICASSP)}.\hskip 1em plus 0.5em minus 0.4em\relax IEEE, 2018, pp.
  4904--4908.

\bibitem{das2015tl}
A.~Das and M.~Hasegawa-Johnson, ``Cross-lingual transfer learning during
  supervised training in low resource scenarios,'' in \emph{Sixteenth Annual
  Conference of the International Speech Communication Association}, 2015.

\bibitem{Dalmia2018low-res-asr}
\BIBentryALTinterwordspacing
S.~Dalmia, R.~Sanabria, F.~Metze, and A.~W. Black, ``Sequence-based
  multi-lingual low resource speech recognition,'' \emph{CoRR}, vol.
  abs/1802.07420, 2018. [Online]. Available:
  \url{http://arxiv.org/abs/1802.07420}
\BIBentrySTDinterwordspacing

\bibitem{Tung2020low-res-asr}
V.~T. Pham, H.~Xu, K.~Yerbolat, Z.~Zheng, E.~S. Chng, C.~Ni, B.~Ma, and H.~Li,
  ``Independent language model architecture for end-to-end asr,'' in \emph{2020
  IEEE International Conference on Acoustics, Speech and Signal Processing},
  2020.

\bibitem{itoh2012data-selection}
N.~Itoh, T.~N. Sainath, D.~N. Jiang, J.~Zhou, and B.~Ramabhadran, ``N-best
  entropy based data selection for acoustic modeling,'' in \emph{2012 IEEE
  International Conference on Acoustics, Speech and Signal Processing
  (ICASSP)}.\hskip 1em plus 0.5em minus 0.4em\relax IEEE, 2012, pp. 4133--4136.

\bibitem{wei2014data-selection-lvcsr}
K.~Wei, Y.~Liu, K.~Kirchhoff, C.~Bartels, and J.~Bilmes, ``Submodular subset
  selection for large-scale speech training data,'' in \emph{2014 IEEE
  International Conference on Acoustics, Speech and Signal Processing
  (ICASSP)}.\hskip 1em plus 0.5em minus 0.4em\relax IEEE, 2014, pp. 3311--3315.

\bibitem{wei2014data-selection-unsup}
K.~Wei, Y.~Liu, K.~Kirchhoff, and J.~Bilmes, ``Unsupervised submodular subset
  selection for speech data,'' in \emph{2014 IEEE International Conference on
  Acoustics, Speech and Signal Processing (ICASSP)}.\hskip 1em plus 0.5em minus
  0.4em\relax IEEE, 2014, pp. 4107--4111.

\bibitem{chen2015prn-lex}
G.~Chen, H.~Xu, M.~Wu, D.~Povey, and S.~Khudanpur, ``Pronunciation and silence
  probability modeling for asr,'' in \emph{Sixteenth Annual Conference of the
  International Speech Communication Association}, 2015.

\bibitem{albert2018e2e-libri}
A.~Zeyer, K.~Iriel, R.~Schl\"{u}ter, and N.~Hermann, ``Improved training of
  end-to-end attention models for speech recognition,'' in \emph{Proc. of
  INTERSPEECH}, 2018.

\bibitem{povey2018tdnnf}
D.~Povey, G.~Cheng, Y.~Wang, K.~Li, H.~Xu, M.~Yarmohamadi, and S.~Khudanpur,
  ``Semi-orthogonal low-rank matrix factorization for deep neural networks,''
  in \emph{Proceedings of INTERSPEECH}, 2018.

\bibitem{zhou2018comparison-e2e}
S.~Zhou, L.~Dong, S.~Xu, and B.~Xu, ``A comparison of modeling units in
  sequence-to-sequence speech recognition with the transformer on {Mandarin
  Chinese},'' in \emph{International Conference on Neural Information
  Processing}.\hskip 1em plus 0.5em minus 0.4em\relax Springer, 2018, pp.
  210--220.

\bibitem{karita2019comparative-e2e}
S.~Karita, N.~Chen, T.~Hayashi, T.~Hori, H.~Inaguma, Z.~Jiang, M.~Someki,
  N.~E.~Y. Soplin, R.~Yamamoto, X.~Wang \emph{et~al.}, ``A comparative study on
  transformer vs rnn in speech applications,'' \emph{arXiv preprint
  arXiv:1909.06317}, 2019.

\bibitem{povey2016lf-mmi}
D.~Povey, V.~Peddinti, D.~Galvez, P.~Ghahremani, V.~Manohar, X.~Na, Y.~Wang,
  and S.~Khudanpur, ``Purely sequence-trained neural networks for asr based on
  lattice-free mmi.'' in \emph{Interspeech}, 2016, pp. 2751--2755.

\bibitem{saon2013ivector}
G.~Saon, H.~Soltau, D.~Nahamoo, and M.~Picheny, ``Speaker adaptation of neural
  network acoustic models using i-vectors,'' in \emph{2013 IEEE Workshop on
  Automatic Speech Recognition and Understanding}.\hskip 1em plus 0.5em minus
  0.4em\relax IEEE, 2013, pp. 55--59.

\bibitem{peddinti2015tdnn}
V.~Peddinti, D.~Povey, and S.~Khudanpur, ``A time delay neural network
  architecture for efficient modeling of long temporal contexts,'' in
  \emph{Sixteenth Annual Conference of the International Speech Communication
  Association}, 2015.

\bibitem{dahl2013relu}
G.~E. Dahl, T.~N. Sainath, and G.~E. Hinton, ``Improving deep neural networks
  for lvcsr using rectified linear units and dropout,'' in \emph{2013 IEEE
  international conference on acoustics, speech and signal processing}.\hskip
  1em plus 0.5em minus 0.4em\relax IEEE, 2013, pp. 8609--8613.

\bibitem{ko2015da}
T.~Ko, V.~Peddinti, D.~Povey, and S.~Khudanpur, ``Audio augmentation for speech
  recognition,'' in \emph{Sixteenth Annual Conference of the International
  Speech Communication Association}, 2015.

\bibitem{duc2019grapheme}
D.~Le, X.~Zhang, W.~Zheng, C.~F{\"{u}}gen, G.~Zweig, and M.~L. Seltzer, ``From
  senones to chenones: Tied context-dependent graphemes for hybrid speech
  recognition,'' in \emph{{IEEE} Automatic Speech Recognition and Understanding
  Workshop, {ASRU} 2019, Singapore, December 14-18, 2019}, 2019.

\end{thebibliography}
\end{document}